\documentclass{llncs}

\usepackage{graphicx,amsmath,amssymb,url,subfigure}
\usepackage[ruled]{algorithm2e}
\usepackage{multirow,setspace,multitoc}

\renewcommand{\leq}{\leqslant}

\begin{document}
\title{Revealing the Autonomous System Taxonomy: \\
The Machine Learning Approach}

\author{ Xenofontas Dimitropoulos\inst{1} \and Dmitri Krioukov\inst{2}
        \and George Riley\inst{1} \and KC Claffy\inst{2}}
\institute{Georgia Institute of Technology\\
        \email{fontas@ece.gatech.edu}\\
        \email{riley@ece.gatech.edu}
\and
  Cooperative Association for Internet Data Analysis (CAIDA)\\
  \email{dima@caida.org}\\
  \email{kc@caida.org}
}

\maketitle

\begin{abstract}

Although the Internet AS-level topology has been extensively
studied over the past few years, little is known about the
details of the AS taxonomy.  An AS "node" can represent a wide
variety of organizations, e.g., large ISP, or small private
business, university, with vastly different network characteristics,
external connectivity patterns, network growth tendencies, and
other properties that we can hardly neglect while working on veracious
Internet representations in simulation environments. In this
paper, we introduce a radically new approach based on
machine learning techniques to map all the ASes in the Internet
into a natural AS taxonomy.
We successfully classify~95.3\% of ASes with expected accuracy
of~78.1\%.
We release to the community the
AS-level topology dataset augmented with: 1)~the AS taxonomy
information and 2)~the set of AS attributes we used to classify
ASes.
We believe that this dataset will serve as an invaluable addition
to further understanding of the structure and evolution of the
Internet.
\end{abstract}

\section{Introduction}

The rapid expansion of the Internet in the last two decades has produced
a large-scale system of thousands of diverse, independently managed
networks that collectively provide global connectivity across a wide
spectrum of geopolitical environments. From 1997 to 2005 the number of
globally routable AS identifiers has increased from less than 2,000 to more
than 20,000, exerting significant pressure on interdomain routing as
well as other functional and structural parts of the Internet. This
impressive growth has resulted in a heterogenous and highly complex
system that challenges accurate and realistic modeling of the
Internet infrastructure. In particular, the AS-level topology
is an intermix of networks owned and operated by many different
organizations, e.g., backbone providers, regional providers,
access providers, universities and private companies. Statistical
information that faithfully characterizes different AS types
is on the critical path toward understanding the structure of the
Internet, as well as for modeling its topology and growth.

In topology modeling, knowledge of AS types is mandatory for augmenting
synthetically constructed or measured AS topologies with realistic
intra-AS and inter-AS router-level topologies. For example, we expect
the network of a dual-homed university to be drastically different from
that of a dual-homed small company. The university will likely contain
dozens of internal routers, thousands of hosts, and many other network
elements (switches, servers, firewalls). On the other hand, the small
company will most probably have a single router and a simple network
topology. Since there is such a diversity among different network types,
we cannot accurately augment the AS-level topology with appropriate
router-level topologies if we cannot characterize the composing ASes.

Moreover, annotating the ASes in the AS topology with their types is a
prerequisite for modeling the evolution of the Internet, since different
types of ASes exhibit different growth patterns. For example, Internet
Service Providers (ISP) grow by attracting new customers and by engaging
in business agreements with other ISPs. On the other hand, small companies
that connect to the Internet through one or few ISPs do not grow
significantly over time.  Thus, categorizing different types of ASes in
the Internet is necessary to identify network evolution patterns and
develop accurate evolution models.

An AS taxonomy is also necessary for mapping IP addresses to different types
of users. For example, in traffic analysis studies its often required to
distinguish between packets that come from home and business users.
Given an AS taxonomy, its possible to realize this goal by checking the
type of AS that originates the prefix in which an IP address lies.

In this work, we introduce a radically new approach based on machine
learning to construct a representative AS taxonomy. We develop an algorithm
to classify ASes based on empirically observed differences between AS
characteristics.  We use a large set of data from the Internet Routing
Registries~(IRR)~\cite{irr} and from RouteViews~\cite{routeviews} to
identify intrinsic differences between ASes of different types. Then,
we employ a novel machine learning technique to build a
classification algorithm that exploits these differences to classify
ASes into six representative classes that reflect ASes with different
network properties and infrastructures. We derive macroscopic statistics
on the different types of ASes in the Internet and validate our results
using a sample of~1200 manually identified AS types.
Our validation demonstrates that our classification algorithm achieves
high accuracy:~78.1\% of the examined classifications were correct.
Finally, we make our results and our classifier publicly available
to promote further research and understanding of the Internet's structure
and evolution.

In Section~\ref{sec:rel} we start with a brief discussion of related work.
Section~\ref{sec:data} describes the data we used, and
in Section~\ref{sec:classes} we specify the set of AS classes we use in
our experiments.  Section~\ref{sec:clas} introduces our classification
approach and results. We validate them in Section~\ref{sec:val} and
conclude in Section~\ref{sec:con}.

\section{Related Work}
\label{sec:rel}

Several works have developed techniques decomposing the AS topology into
different levels or tiers based on connectivity properties of BGP-derived
AS graphs. Govindan and Reddy~\cite{GoRe97} propose a classification of
ASes into four levels based on their AS degree. Ge {\it et
al}.~\cite{GeFiJaGa01} classify ASes into seven tiers based on inferred
customer-to-provider relationships. Their classification
exploits the idea that provider ASes should be in higher tiers than their
customers. Subramanian {\it et al}.~\cite{SuAgReKa02} classify ASes into
five tiers based on inferred customer-to-provider as well as peer-to-peer
relationships.

Our work differs from previous approaches in the following ways:
\begin{enumerate}
\item We do not employ heuristics and ad-hoc thresholds to define the AS levels.
Instead, we use a novel machine learning algorithm to identify intrinsic
features distinguishing different AS types.
\item We do not rely exclusively on AS graphs, which often miss a substantial fraction of the
true AS links in the Internet, resulting in incomplete AS topologies. Instead, we use an extensive
set of diverse data including IRR records, inferred AS relationships, IP prefixes, and AS graphs.
\item We do not classify ASes into hierarchies of levels or tiers extracted from AS graphs using
degree-based or more sophisticated mechanisms; these methods tend to mix ASes with
substantially different network properties into a single AS group.\footnote{According
to our analysis, small regional providers often have small AS degrees,
as low as~1 or~2. The previous heuristics thus tend to assign these ASes to the lowest levels, where
small companies and multihomed customers naturally reside.} Instead, we specify a representative set
of AS classes characterized by unique signatures of network properties.
\end{enumerate}

\section{Data Sources and AS Attributes}
\label{sec:data}

To construct the set of AS attributes that we use in our AS classification, we
collect data from the following databases and measurement projects:

1)~{\bf IRRs}~\cite{irr}.  The IRRs constitute a distributed database
containing records on ASes' routing policies, assigned IP prefixes,
contact information, etc.  A natural approach to identifying the type
of an AS, given its AS number, is to lookup the AS number in the IRRs
and examine its organization description record. In the RPSL~\cite{rpsl}
terminology, this record is the {\tt descr} attribute of the RPSL
class {\tt aut-num}. It contains the name or a short description of the
organization that owns the AS number.  For example, the following are
entries for the {\tt descr} attribute found in the IRRs: ``Intervivo
Networks, a broadband Internet access provider'' and ``Auckland Peering
Exchange''. The {\tt descr} attribute does not have a standard representation.
It usually consists of a short description as in the examples above, but
in some cases it only contains an acronym, e.g., ``KPMG LLP'', ``LTI''.
For the purposes of this work, the {\tt descr} record is our first AS
attribute, from which we extract useful information by means of text
classification techniques.  We downloaded the mapping of AS numbers to
organization description records on 04/08/2005 from the CIDR
Report~\cite{CidrReport}, which provides on a daily basis mappings
of AS numbers to organization description records extracted from ARIN,
RIPE NCC, LAPNIC, APNIC, KRNIC, TWNIC, and JPNIC databases. We preprocess
the organization records by removing stop words, i.e., words with little
semantic meaning, such as ``of'' or ``the'', using the stop word
list~\cite{StopList}.  Then, using the Porter stemming
algorithm~\cite{Po80}, we replace words with their stem.

We note that IRRs contain significant portions of incomplete or
obsolete records, which is not a serious problem for this study since
we are only concerned with the {\tt descr} attribute, which
changes relatively rarely.

2)~{\bf RouteViews}~\cite{routeviews}. RouteViews is a measurement project
that collects and archives a union of BGP tables from a large number of ISPs.
We download all 12 BGP table snapshots archived from the collector {\tt
route-views2.oregon-ix.net} on 07/18/2005. For each table snapshot we
extract AS paths and remove AS sets and private AS numbers. Then, we
merge the extracted AS paths into an AS topology and use the AS relationship
inference heuristics of~\cite{DiKrHuClRi05,DiKrFoHuClRi05} to annotate
the AS links with customer-to-provider and peer-to-peer relationships.
Having the AS relationships inferred, we calculate the following three
attributes for every AS: the number of providers, the number of customers,
and the number of peers a given AS is connected to.  Large ISPs typically
have a large number of customers, zero providers, and a small number of
peers, while small ISPs typically have few providers, a small number of
customers and a large number of peers. Stub university or company networks
typically have zero customers, zero or few peers and a small number of
providers.

From the RouteViews data, we also extract information on IP prefixes.
We use the chronologically first table snapshot from the same snapshot
set to construct a mapping of AS numbers to the IP prefixes they advertise.
Then, for each AS we count the number of advertised prefixes and use
this number as another AS attribute. Small ASes, with tiny portions of
IP address space allocated to them, as well as older ASes with large IP
blocks, tend to advertise few prefixes. On the contrary, large ASes with
relatively high IP address utilization and diversified routing policies
tend to advertise many prefixes of various lengths.

One problem with this attribute is that IP prefixes are of drastically
different sizes. Advertised IPv4 prefixes range in size from a~/8,
covering~$2^{24}$ IP addresses, down to a~/32, covering just one address.
The prefix length of 24 bits is generally the smallest IPv4 prefix size
that is globally routed, which suggests our last AS attribute to be the
number of unique~/24 prefixes found within the union of address space
advertised by the AS. This attribute is likely to have small values for
smaller ASes that use and advertise smaller portions of IP address space,
while it is likely to be at its maximum for large or old ASes (those that
appeared in the Internet early, e.g., some military and academic networks)
since they often have huge chunks of assigned IP address space that they
utilize scarcely.

In summary, we collect data from the IRRs and RouteViews.
We find~19,537 ASes. Using the collected data, we annotate every AS with the
following six attributes: 1)~the organization description record (the {\em description\/} attribute),
2)~the number of inferred customers (the {\em customer\/} attribute), 3)~the number of inferred
providers (the {\em provider\/} attribute), 4)~the number of inferred peers (the {\em peer\/} attribute),
5)~the number of advertised IP prefixes (the {\em prefix\/} attribute), and 6)~the equivalent number
of~/24 prefixes covering all the advertised IP space (the {\em space\/} attribute).

\section{The AS Class Set}
\label{sec:classes}

In this work, we focus on network properties of an AS as the main criterion determining the set of AS classes.
In other words, to construct the AS class set, we use the rule that ASes in the same class should have similar
network properties, while ASes in different classes should have different network properties.
ASes in the same class may still have significant network differences, however these differences should be small
compared with differences between networks of different classes. For example, a small university and a large
university may have quite different networks, however these differences are less significant compared to the
differences between a typical university network and a typical ISP network.

Besides employing our {\it de facto} empirical knowledge, we perform the following experiment to specify
the set of AS classes. We {\it randomly} select~1200 ASes and then, for each AS, we examine its attributes,
visit its website (if possible), search for references to its organization name and determine its business
profile.
After examining the spectrum of these~1200 ASes, we construct our AS class set:
\begin{enumerate}
\item {\em Large ISPs}: Large backbone providers, tier-1 ISPs, with intercontinental networks.
\item {\em Small ISPs}:  Regional and access providers with small metropolitan or larger regional networks.
\item {\em Customer ASes}: Companies or organizations that run their own networks but as opposed to members
of the previous two classes do not provide Internet connectivity services. We find a wide range of ASes in this
class, like web hosting companies, technology companies, consulting companies, hospitals, banks, military networks,
government networks, etc.
\item {\em Universities}: University or college networks. We distinguish these networks from members of the Customer
AS class, since they typically have substantially larger networks that serve thousands of end hosts.
\item {\em Internet exchange points (IXPs)}: Small networks serving as interconnection points for the members of the first two classes.
\item {\em Network information centers (NICs)}: Networks that host important network infrastructure, such as root or TLD servers.
\end{enumerate}

\section{AS Classification: Algorithms and Results}
\label{sec:clas}

We build our classification algorithm using {\em AdaBoost}~\cite{AdaBoost}, a very powerful machine learning technique.
The main idea behind AdaBoost is to combine multiple simple classification rules into a efficient composite classifier.
These simple classification rules are called {\em weak hypotheses} and by definition are only required to perform
slightly better than random guessing. Intuitively, weak hypotheses reflect simple ``rules of thumb'' that are
usually much easier to construct than a complex classifier. AdaBoost works iteratively
over a set of training examples; at each iteration it finds a weak hypothesis that performs
well on the examples which the weak hypotheses of previous iterations erroneously classified. One constructs
a weak hypothesis by means of a {\em weak learning algorithm} or simply {\em weak learner}. The power of AdaBoost lies in a
well-developed theoretical framework that intelligently combines the weak hypotheses into a composite classifier.

Let~$X$ denote the set of ASes that we want to classify and~$Y$ be the set of possible classes, such that each AS~$x \in X$
belongs to a unique class in~$Y$. If~$y \in Y$ is the correct class for AS~$x$, than let $P(x,y) = 1$, otherwise~$P(x,y) = -1$.
The goal is to find a classifier that for each AS produces a {\it ranking} of all the possible classes. More formally, we
will compute a ranking function~$f:X \times Y \rightarrow \mathbb{R}$: for each AS~$x \in X$, the classes in~$Y$ will be
ordered according to~$f(x,\cdot)$---the higher the value of~$f(x,y)$, the more likely~$x$ belongs to~$y$.

\begin{algorithm}[t]

\caption{{\it AdaBoost.MH} pseudocode\label{fig:ada}}
\dontprintsemicolon
\SetLine
\SetKwInput{KwInit}{Initialize}
\KwIn{$S= \{(x_1,y_1),\ldots,(x_m,y_m)\}$}
\KwInit{$D_1(x,y)=1/mk$\tcp*{$k$ is the total number of classes}}
\For{ $t = 1$ \KwTo $T$}{
  Pass distribution~$D_t$ and examples~$S$ to weak learner\;
  Get weak hypothesis~$h_t:X \times Y \rightarrow \mathbb{R}$\;
  Update distribution \[D_{t+1}(x,y) = \frac{ D_t(x,y) exp(- P(x,y) h_t(x,y) ) }{Z_t}\] \;
\tcp{where $Z_t$ is a normalization coefficient}\;
\tcp{chosen so that $D_{t+1}$ will be a distribution}\;
}
\KwOut{$f(x,y) = \sum_{t=1}^T{h_t(x,y)}$}
\end{algorithm}
\setcounter{figure}{1}

In Figure~\ref{fig:ada}, we illustrate the AdaBoost.MH algorithm~\cite{Boostexter}, a special member of the AdaBoost algorithm
family that is suited for solving multiclass problems.
Let $S= \{(x_1,y_1),\ldots,(x_m,y_m)\}$ be the set of training examples. In our case, we construct~$S$ by manually
determining the correct class of~1220 ASes: 1200 are randomly selected; and 20 are well-known large ISPs and IXPs, which we
use to increase the number of these two types of ASes in the initial random sample.
Let~$T$ be the total number of
iterations. For each iteration $t = 1 \ldots T$, we maintain a distribution~$D_t$ of weights over the set of examples
and classes $D_t:X \times Y \rightarrow \mathbb{R}$. At the first iteration, we initialize $D_t$ to
the uniform distribution, i.e., $D_1$ is constant for all $(x_i,y_i)$, $1 \leq i \leq m$ .
At each subsequent iteration, we pass the distribution~$D_t$ and the training examples~$S$
to a weak learner that computes a weak hypothesis~$h_t:X \times Y \rightarrow \mathbb{R}$. A positive (negative) sign of the
weak hypothesis~$h_t(x,y)$ corresponds to the prediction that AS~$x$ is (is not) a member of class~$y$.
The value $|h_t(x,y)|$ of the weak hypothesis reflects the confidence level of the prediction.
Then, we update the distribution~$D_t$ so that the~$x$,~$y$ pairs that were erroneously
predicted, i.e., the signs of~$h(x,y)$ and~$P(x,y)$ differ, receive a exponentially higher weight. By assigning higher weight to
incorrect predictions, we force the algorithm to focus on these difficult examples in the next round. The final classifier~$f$ is
the sum of votes of the weak hypotheses in all rounds~$h_t$, $t = 1 \ldots T$.

A weak hypothesis is equivalent to a one-level decision tree that checks a single AS attribute.
For the description AS attribute, a weak hypothesis searches for the presence of a term or a sequence of terms in
a given record, and if a match occurs, it outputs a confidence value for each of the classes. For example, upon finding the term
``university'' in the record ``Seoul National University of Education'' the weak hypothesis will likely output a high positive
confidence value for the University AS class and a negative confidence value for the other AS classes. For scalar attributes, a weak
hypothesis asks if a given attribute value is above or below a certain threshold. Depending on the outcome, the hypothesis outputs
different confidence values.

The weak learner builds a weak hypothesis by exhaustively evaluating the attributes in the given weighted training examples.
For a text attribute, it builds a candidate weak hypothesis by evaluating all possible terms and sequences of terms. For each
term or sequence of terms, it calculates the appropriate confidence values by minimizing the Hamming loss, which is the
fraction of examples~$x$ and classes~$y$, for which the sign of the final classifier~$f(x,y)$ differs from~$P(x,y)$.
Similarly, for each scalar attribute, the weak learner builds a candidate weak hypothesis by exhaustively
searching the threshold and confidence values minimizing the Hamming loss. On its output, the learner returns the weak
hypothesis attaining the minimum Hamming loss.\footnote{See~\cite{Boostexter} for the analytic
expression of the Hamming loss.}

\setlength{\unitlength}{1mm}

\begin{table}[t]
\centering
\caption{The list of weak hypotheses computed by AdaBoost.MH in the first six iterations. The first column is
the iteration number; the second is the AS attribute that the weak hypothesis is checking; the third is the
term or threshold of the weak hypothesis; and the remaining columns depict the computed negative or positive
confidence values for each of the AS classes.}

\label{table}

\begin{tabular}{ccc|cccccc}
Round & Attribute & Term/Threshold & L.ISP & S.ISP & Cusmr & Uni & IXP & NIC\\
\hline

\multirow{2}{*}{1} & \multirow{2}{*}{space} & $< 8.5$
&

\begin{picture}(1,7.2)
\put(1,-0.0163121304999998){\framebox(1,3.5163121305){}}
\linethickness{.8pt}
\put(0,3.5){\line(1,0){3}}
\end{picture}

&

\begin{picture}(1,7.2)
\put(1,2.4735347919){\framebox(1,1.0264652081){}}
\linethickness{.8pt}
\put(0,3.5){\line(1,0){3}}
\end{picture}

&

\begin{picture}(1,7.2)
\put(1,3.5){\framebox(1,0.7797397768){}}
\linethickness{.8pt}
\put(0,3.5){\line(1,0){3}}
\end{picture}

&

\begin{picture}(1,7.2)
\put(1,1.8325260019){\framebox(1,1.6674739981){}}
\linethickness{.8pt}
\put(0,3.5){\line(1,0){3}}
\end{picture}

&

\begin{picture}(1,7.2)
\put(1,1.4639146476){\framebox(1,2.0360853524){}}
\linethickness{.8pt}
\put(0,3.5){\line(1,0){3}}
\end{picture}

&

\begin{picture}(1,7.2)
\put(1,1.2714864672){\framebox(1,2.2285135328){}}
\linethickness{.8pt}
\put(0,3.5){\line(1,0){3}}
\end{picture}

\\

& & $> 8.5$

&

\begin{picture}(1,7.2)
\put(1,1.2708921135){\framebox(1,2.2291078865){}}
\linethickness{.8pt}
\put(0,3.5){\line(1,0){3}}
\end{picture}

&

\begin{picture}(1,7.2)
\put(1,3.5){\framebox(1,0.0336004254){}}
\linethickness{.8pt}
\put(0,3.5){\line(1,0){3}}
\end{picture}

&

\begin{picture}(1,7.2)
\put(1,3.1608508751){\framebox(1,0.3391491249){}}
\linethickness{.8pt}
\put(0,3.5){\line(1,0){3}}
\end{picture}

&

\begin{picture}(1,7.2)
\put(1,2.2256887679){\framebox(1,1.2743112321){}}
\linethickness{.8pt}
\put(0,3.5){\line(1,0){3}}
\end{picture}

&

\begin{picture}(1,7.2)
\put(1,1.5301809138){\framebox(1,1.9698190862){}}
\linethickness{.8pt}
\put(0,3.5){\line(1,0){3}}
\end{picture}

&

\begin{picture}(1,7.2)
\put(1,1.990387677){\framebox(1,1.509612323){}}
\linethickness{.8pt}
\put(0,3.5){\line(1,0){3}}
\end{picture}

\\ \hline

2 & description & ``network inform''

&

\begin{picture}(1,7.2)
\put(1,2.2216833816){\framebox(1,1.2783166184){}}
\linethickness{.8pt}
\put(0,3.5){\line(1,0){3}}
\end{picture}

&

\begin{picture}(1,7.2)
\put(1,1.1210242844){\framebox(1,2.3789757156){}}
\linethickness{.8pt}
\put(0,3.5){\line(1,0){3}}
\end{picture}

&

\begin{picture}(1,7.2)
\put(1,1.1349191402){\framebox(1,2.3650808598){}}
\linethickness{.8pt}
\put(0,3.5){\line(1,0){3}}
\end{picture}

&

\begin{picture}(1,7.2)
\put(1,1.7399693498){\framebox(1,1.7600306502){}}
\linethickness{.8pt}
\put(0,3.5){\line(1,0){3}}
\end{picture}

&

\begin{picture}(1,7.2)
\put(1,2.0556227578){\framebox(1,1.4443772422){}}
\linethickness{.8pt}
\put(0,3.5){\line(1,0){3}}
\end{picture}

&

\begin{picture}(1,7.2)
\put(1,3.5){\framebox(1,3.2284513535){}}
\linethickness{.8pt}
\put(0,3.5){\line(1,0){3}}
\end{picture}

\\ \hline

\multirow{2}{*}{3} & \multirow{2}{*}{customer} & $< 1.5$
&

\begin{picture}(1,7.2)
\put(1,0.6010428496){\framebox(1,2.8989571504){}}
\linethickness{.8pt}
\put(0,3.5){\line(1,0){3}}
\end{picture}

&

\begin{picture}(1,7.2)
\put(1,3.4404276559){\framebox(1,0.0595723441){}}
\linethickness{.8pt}
\put(0,3.5){\line(1,0){3}}
\end{picture}

&

\begin{picture}(1,7.2)
\put(1,3.5){\framebox(1,0.1111626141){}}
\linethickness{.8pt}
\put(0,3.5){\line(1,0){3}}
\end{picture}

&

\begin{picture}(1,7.2)
\put(1,3.5){\framebox(1,0.0390868724){}}
\linethickness{.8pt}
\put(0,3.5){\line(1,0){3}}
\end{picture}

&

\begin{picture}(1,7.2)
\put(1,3.4052849295){\framebox(1,0.0947150705){}}
\linethickness{.8pt}
\put(0,3.5){\line(1,0){3}}
\end{picture}

&

\begin{picture}(1,7.2)
\put(1,2.7965965279){\framebox(1,0.7034034721){}}
\linethickness{.8pt}
\put(0,3.5){\line(1,0){3}}
\end{picture}

\\

& & $> 1.5$

&

\begin{picture}(1,7.2)
\put(1,3.5){\framebox(1,0.8634883271){}}
\linethickness{.8pt}
\put(0,3.5){\line(1,0){3}}
\end{picture}

&

\begin{picture}(1,7.2)
\put(1,3.5){\framebox(1,0.7049122965){}}
\linethickness{.8pt}
\put(0,3.5){\line(1,0){3}}
\end{picture}

&

\begin{picture}(1,7.2)
\put(1,2.3929064144){\framebox(1,1.1070935856){}}
\linethickness{.8pt}
\put(0,3.5){\line(1,0){3}}
\end{picture}

&

\begin{picture}(1,7.2)
\put(1,3.1135206426){\framebox(1,0.3864793574){}}
\linethickness{.8pt}
\put(0,3.5){\line(1,0){3}}
\end{picture}

&

\begin{picture}(1,7.2)
\put(1,3.5){\framebox(1,0.4135959469){}}
\linethickness{.8pt}
\put(0,3.5){\line(1,0){3}}
\end{picture}

&

\begin{picture}(1,7.2)
\put(1,2.7023153932){\framebox(1,0.7976846068){}}
\linethickness{.8pt}
\put(0,3.5){\line(1,0){3}}
\end{picture}

\\ \hline

4 & description & ``exchang''

&

\begin{picture}(1,7.2)
\put(1,2.6649888658){\framebox(1,0.8350111342){}}
\linethickness{.8pt}
\put(0,3.5){\line(1,0){3}}
\end{picture}

&

\begin{picture}(1,7.2)
\put(1,1.2936841338){\framebox(1,2.2063158662){}}
\linethickness{.8pt}
\put(0,3.5){\line(1,0){3}}
\end{picture}

&

\begin{picture}(1,7.2)
\put(1,2.6048529329){\framebox(1,0.8951470671){}}
\linethickness{.8pt}
\put(0,3.5){\line(1,0){3}}
\end{picture}

&

\begin{picture}(1,7.2)
\put(1,1.9378176669){\framebox(1,1.5621823331){}}
\linethickness{.8pt}
\put(0,3.5){\line(1,0){3}}
\end{picture}

&

\begin{picture}(1,7.2)
\put(1,3.5){\framebox(1,2.4512993094){}}
\linethickness{.8pt}
\put(0,3.5){\line(1,0){3}}
\end{picture}

&

\begin{picture}(1,7.2)
\put(1,2.4186244687){\framebox(1,1.0813755313){}}
\linethickness{.8pt}
\put(0,3.5){\line(1,0){3}}
\end{picture}

\\ \hline

5 & description & ``univers''

&

\begin{picture}(1,7.2)
\put(1,2.7987866841){\framebox(1,0.7012133159){}}
\linethickness{.8pt}
\put(0,3.5){\line(1,0){3}}
\end{picture}

&

\begin{picture}(1,7.2)
\put(1,1.8342091922){\framebox(1,1.6657908078){}}
\linethickness{.8pt}
\put(0,3.5){\line(1,0){3}}
\end{picture}

&

\begin{picture}(1,7.2)
\put(1,2.0500111654){\framebox(1,1.4499888346){}}
\linethickness{.8pt}
\put(0,3.5){\line(1,0){3}}
\end{picture}

&

\begin{picture}(1,7.2)
\put(1,3.5){\framebox(1,2.4096956442){}}
\linethickness{.8pt}
\put(0,3.5){\line(1,0){3}}
\end{picture}

&

\begin{picture}(1,7.2)
\put(1,2.1049456495){\framebox(1,1.3950543505){}}
\linethickness{.8pt}
\put(0,3.5){\line(1,0){3}}
\end{picture}

&

\begin{picture}(1,7.2)
\put(1,2.263214415){\framebox(1,1.236785585){}}
\linethickness{.8pt}
\put(0,3.5){\line(1,0){3}}
\end{picture}

\\ \hline

\multirow{2}{*}{6} & \multirow{2}{*}{customer} & $< 97$
&

\begin{picture}(1,7.2)
\put(1,0.9666483892){\framebox(1,2.5333516108){}}
\linethickness{.8pt}
\put(0,3.5){\line(1,0){3}}
\end{picture}

&

\begin{picture}(1,7.2)
\put(1,3.5){\framebox(1,0.0514789400){}}
\linethickness{.8pt}
\put(0,3.5){\line(1,0){3}}
\end{picture}

&

\begin{picture}(1,7.2)
\put(1,3.5){\framebox(1,0.0423646810){}}
\linethickness{.8pt}
\put(0,3.5){\line(1,0){3}}
\end{picture}

&

\begin{picture}(1,7.2)
\put(1,3.3302953706){\framebox(1,0.1697046294){}}
\linethickness{.8pt}
\put(0,3.5){\line(1,0){3}}
\end{picture}

&

\begin{picture}(1,7.2)
\put(1,2.9781279276){\framebox(1,0.5218720724){}}
\linethickness{.8pt}
\put(0,3.5){\line(1,0){3}}
\end{picture}

&

\begin{picture}(1,7.2)
\put(1,3.5){\framebox(1,0.0133931961){}}
\linethickness{.8pt}
\put(0,3.5){\line(1,0){3}}
\end{picture}

\\

& & $> 97$

&

\begin{picture}(1,7.2)
\put(1,3.5){\framebox(1,2.4854406092){}}
\linethickness{.8pt}
\put(0,3.5){\line(1,0){3}}
\end{picture}

&

\begin{picture}(1,7.2)
\put(1,1.3250967327){\framebox(1,2.1749032673){}}
\linethickness{.8pt}
\put(0,3.5){\line(1,0){3}}
\end{picture}

&

\begin{picture}(1,7.2)
\put(1,2.3689860715){\framebox(1,1.1310139285){}}
\linethickness{.8pt}
\put(0,3.5){\line(1,0){3}}
\end{picture}

&

\begin{picture}(1,7.2)
\put(1,2.4639524202){\framebox(1,1.0360475798){}}
\linethickness{.8pt}
\put(0,3.5){\line(1,0){3}}
\end{picture}

&

\begin{picture}(1,7.2)
\put(1,2.4179592278){\framebox(1,1.0820407722){}}
\linethickness{.8pt}
\put(0,3.5){\line(1,0){3}}
\end{picture}

&

\begin{picture}(1,7.2)
\put(1,2.7330272347){\framebox(1,0.7669727653){}}
\linethickness{.8pt}
\put(0,3.5){\line(1,0){3}}
\end{picture}

\\ \hline

\end{tabular}
\end{table}

To realize our classification algorithm, we use BoosTexter~\cite{Boostexter},
a publicly available implementation of AdaBoost. In Table~\ref{table} we depict the weak hypotheses
that our algorithm discovered during its first six iterations. For each weak hypothesis, we illustrate
the selected AS attribute, the term or threshold that is looked for, and the computed confidence values.
The first weak hypothesis deals with the space attribute. If the IP address space advertised by an AS is less than~8.5
(equivalent) /24 prefixes then, the hypothesis assigns a positive confidence value to the Customer AS class and negative
confidence values to the other classes. If the value of the space attribute is above~8.5, the hypothesis assigns
negative or very close to zero confidence values to all the classes, which means that in this case it cannot make a confident
positive prediction. The second weak hypothesis checks the description AS attribute for the presence of term
``network inform''.\footnote{Recall that words have been stemmed.} If it finds one, it assigns a high
positive confidence to the NIC AS class and negative confidence values to the other classes. Note, that
in some cases the weak hypothesis assigns zero confidence values, meaning that it abstains from making
any prediction.

We experimentally fix the number of rounds~$T$ to~28, since subsequent iterations lead to overfitting. Overfitting
is a common problem in machine learning. It is a consequence of too extensive training of an algorithm on one dataset.
The undesired effect is that the algorithm is memorizing the training examples instead of extracting concepts from
them. Fortunately, we can easily detect if the algorithm tends to overfit by examining the produced classification
rules.

Having the number of iterations fixed, we finally apply our classifier to the set of~19,537 ASes and calculate the
ranking of the AS classes for them. For each AS we consider the highest ranked class as its predicted class.
If the class with the highest rank value for an AS has the confidence value less than
or equal to zero, then the classifier abstains from making a prediction since the given information is not sufficient
to produce a reliable assignment. Overall, the classifier abstains from making a prediction for~923 ASes, which
accounts for~4.7\% of the total number of ASes in our dataset.
In Table~\ref{stats} we show the per category classification statistics.
Among the classified ASes, 63.0\% are Customers, 30.1\% are small ISPs, 4.7\% are Universities, 1.8\% are NICs, 0.2\%
are ISPs, and 0.2\% are large IXPs.

\begin{table}
\centering
\caption{Numbers of ASes in each AS class.}
\label{stats}
\begin{tabular}{|c|c|c|c|c|c|c|c|c|}
\hline
&Large ISPs & Small ISPs & Customer ASes & Universities & IXPs & NICs \\ \hline
ASes & 44 & 5,599 & 11,729 & 877  & 33 & 332\\ \hline
\% & 0.2 & 30.1 & 63.0 & 4.7 & 0.2 & 1.8\\
\hline\end{tabular}
\end{table}

\section{Validation}
\label{sec:val}

To validate our results, we employ the standard machine learning methodology called {\em cross-validation}.
Cross-validation is the process of splitting the training examples into two subsets. One then uses the first
subset to train a new classifier and the second subset to validate the results of this new classifier.

From our main set of~1200 training examples, we randomly extract~1100 ASes and use varying size subsets of
these~1100 ASes to train new classifiers. We validate the predictions of the new classifiers against the
remaining~100 examples. We repeat the random selection process~400 times and for each iteration we compute
the following evaluation metrics: 1)~{\em accuracy}, which we define as the percentage of ASes for which
the AS class with the highest rank value is their correct class. The disadvantage of this metric is that it
checks only the top of the ranking, ignoring the remaining positions. To address this problem, we use
2)~{\em coverage}, which we define as the average position number of the correct class of an AS. For each AS,
we number AS class positions incrementally starting from zero for the class with the highest positive
confidence value. Thus, if all the predictions are correct the coverage is zero.

\begin{figure}
\centering
\subfigure[]{
    \label{plot:accuracy}
    \includegraphics[width=2in]{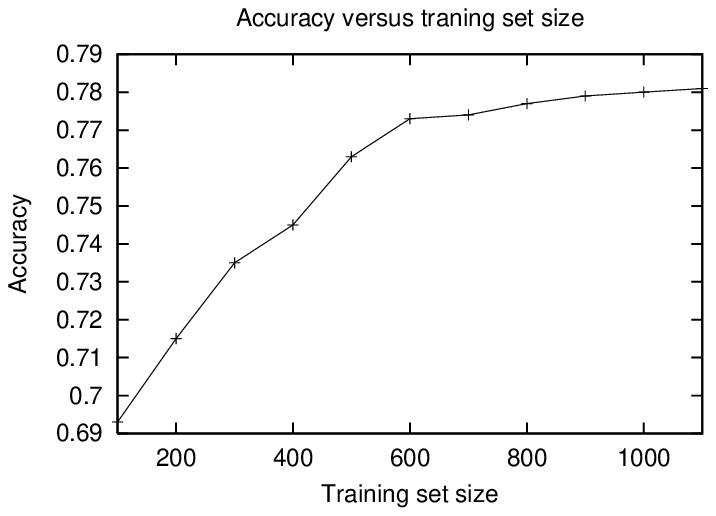}
    }
\subfigure[]{
    \label{plot:coverage}
    \includegraphics[width=2in]{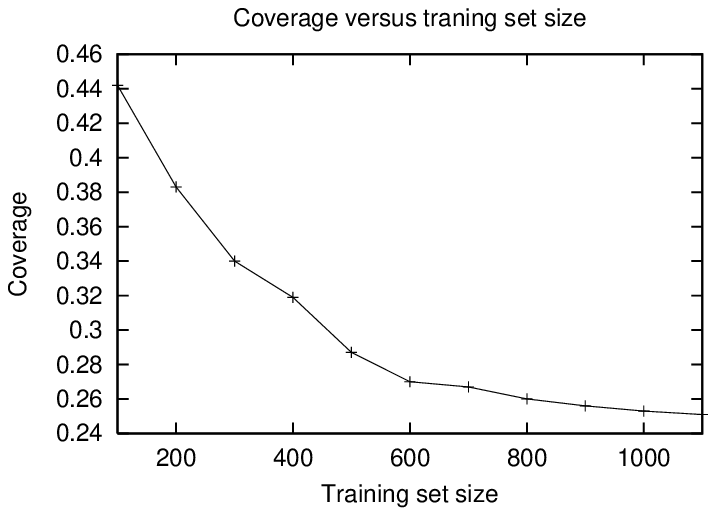}
    }
\caption{Accuracy and coverage of computed predictions versus the size of the training set.}
\end{figure}

In Figures~\ref{plot:accuracy} and~\ref{plot:coverage} we plot the average accuracy and coverage versus
the size of the training set. As the size of the training set increases, the accuracy increases and the
coverage decreases. For~$|S| = 1100$ the accuracy reaches~0.781, e.g., 78.1\%, and the coverage~0.251. The increasing
accuracy trend suggests that for the training size of~1200 that we use in our final classification, the expected accuracy
must be even higher. The low value of the coverage indicates that when the correct class is not of the top rank value,
it is close to the top. More specifically, we find that for~97.7\% of the predictions the correct class is in the top two
positions of the ranking.

We next analyze the per class percentage of correct predictions. We find that for~$|S| = 1100$ the percentage of correct
predictions is on average: 100\% for large ISPs, 100\% for NICs, 100\% for IXPs, 92.8\% for Universities, 79.2\% for
Customer ASes and 72.1\% for small ISPs. The actual distribution of ASes among the classes in our training set is: 684 Customer
ASes, 401 small ISPs, 66 Universities, 36 NICs, 11 IXPs, and 2 large ISPs.
The lower accuracy for customer ASes and small ISPs illustrates that these are the hardest classes to identify.
We explain this effect by similarities between the characteristics of these two AS classes:
1)~more than a half of the small ISPs appear to have the AS degree of~1 or~2, which is also
typical for customer ASes; 2)~some customer ASes, especially web hosting companies, advertise
large numbers of different IP prefixes or large chunks of address space, which is also typical
for ISPs.

In summary, we find that in the examined examples our classifier almost perfectly identifies large ISPs, NICs,
IXPs and universities, while it also produces accurate predictions for customer ASes and small ISPs, which are
the hardest to classify.

\section{Conclusion}
\label{sec:con}

In this work, we establish a classification of ASes required for expanding our understanding of the Internet
infrastructure and for creating realistic models of its topology and evolution. We develop a
novel classification methodology that we apply to an exhaustive set of AS data to obtain
the first statistics on the different AS classes in the Internet. We validate our results and
demonstrate that our classifier achieves accuracy of~78.1\% in the examined data.
To promote further analysis and to inspire development of better topology models, we release to
the community our classification dataset along with the AS class predictions~\cite{AsTaxonomyData}.
To the best of our knowledge, our dataset is the most comprehensive collection of AS macroscopic
characteristics. In addition to AS topology and taxonomy information, it includes organization
description records, AS business relationship information, and information on advertised IP
prefixes and space.

\begin{spacing}{.8}

\end{spacing}
\end{document}